\documentclass[%
 reprint,
superscriptaddress,
nofootinbib,
 amsmath,amssymb,
prl
longbibliography
]{revtex4-2}

\usepackage{makecell} 
\usepackage[colorinlistoftodos]{todonotes} 
\usepackage{natbib}
\usepackage{mathtools}
\usepackage{subfiles}
\usepackage{float}
\usepackage{graphicx}
\usepackage{dcolumn}
\usepackage{bm}
\usepackage{multirow}
\usepackage{appendix}
\usepackage{lipsum} 

\usepackage{amsmath}
\let\oldAA\AA
\renewcommand{\AA}{\text{\normalfont\oldAA}}

\newcommand{\bl}{\mathbf{l}}
\newcommand{\bk}{\mathbf{k}}
\newcommand{\br}{\mathbf{r}}

\newcommand{\bL}{\mathbf{L}}
\newcommand{\bK}{\mathbf{K}}

\newcommand{\ks}{{\mathbf{k}_s}}
\newcommand{\kp}{{\mathbf{k}_p}}

\begin{document}

\title{Deep Learning Sheds Light on Integer and Fractional Topological Insulators}
\author{Xiang Li}
\email{lixiang.62770689@bytedance.com}
\affiliation{ByteDance Research}
\author{Yixiao Chen}
\affiliation{ByteDance Research}
\author{Bohao Li}
\affiliation{School of Physics and Technology,  Wuhan University}

\author{Haoxiang Chen}
\affiliation{ByteDance Research}
\affiliation{School of Physics, Peking University}

\author{Fengcheng Wu}
\email{wufcheng@whu.edu.cn}
\affiliation{School of Physics and Technology, Wuhan University}

\author{Ji Chen}
\email{ji.chen@pku.edu.cn}
\affiliation{School of Physics, Peking University}

\author{Weiluo Ren}
\email{renweiluo@bytedance.com}
\affiliation{ByteDance Research}

\date{\today}
\begin{abstract}
Electronic topological phases of matter, 
characterized by robust boundary states derived from topologically nontrivial bulk states,
are pivotal
for next-generation electronic devices. However, understanding their complex quantum phases, especially at larger scales and fractional fillings with strong electron correlations, 
has long posed a formidable computational challenge.
Here, we employ a deep learning framework to express the many-body wavefunction of topological states in twisted ${\rm MoTe_2}$ systems, where diverse topological states are observed.
Leveraging neural networks, we demonstrate the ability to identify and characterize topological phases, including the integer and fractional Chern insulators as well as the $Z_2$ topological insulators.
Our deep learning approach significantly outperforms traditional methods, not only in computational efficiency but also in accuracy, enabling us to study larger systems and differentiate between competing phases such as fractional Chern insulators and charge density waves. Our predictions align closely with experimental observations, highlighting the potential of deep learning techniques to explore the rich landscape of topological and strongly correlated phenomena.
\end{abstract}

\maketitle

\section{Introduction}

\begin{figure*}[t]
\centering
\includegraphics[width=0.95\textwidth]{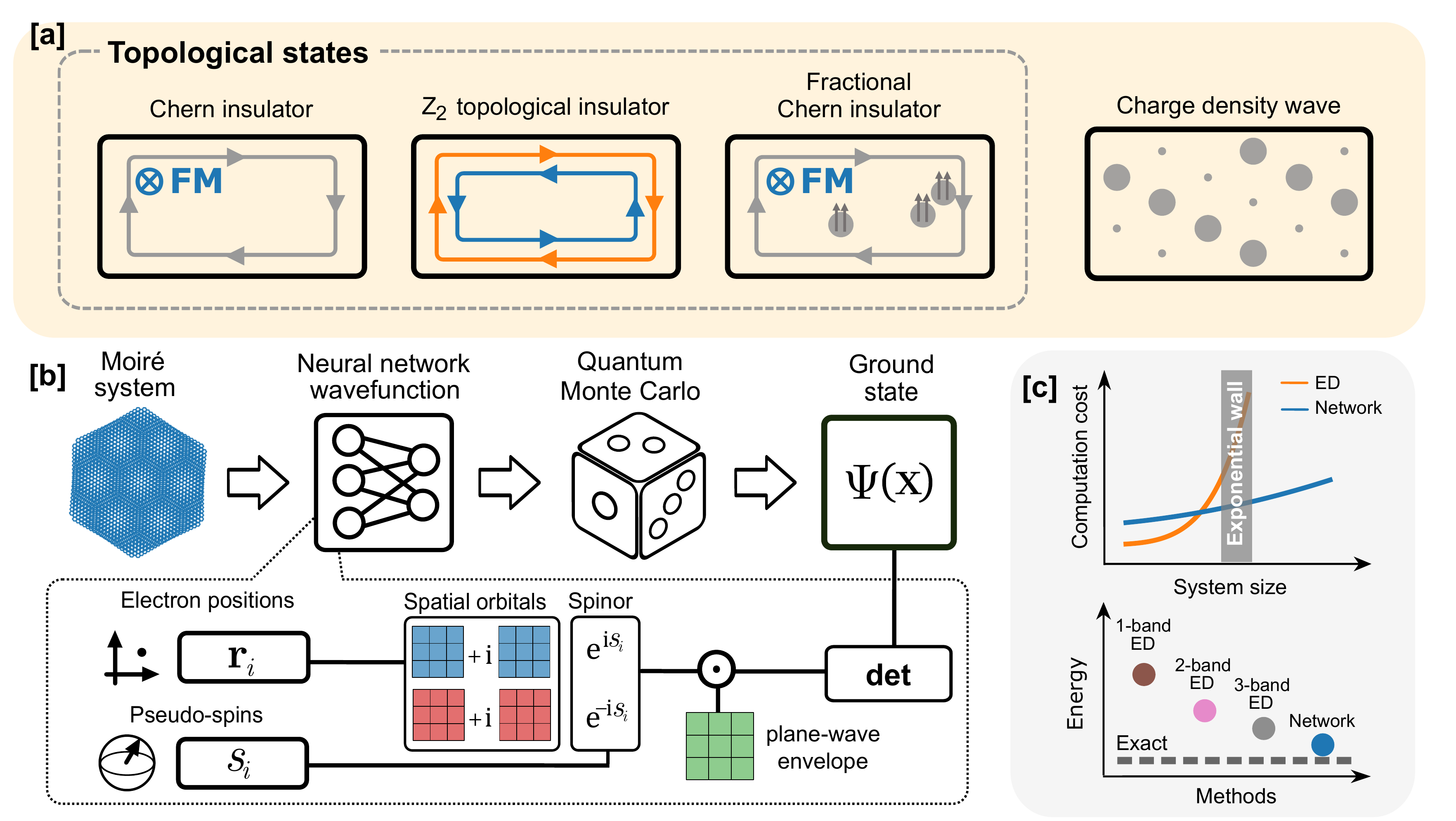}
\caption{\label{fig:concept} \textbf{Sketch of deep learning simulation of topological states.} 
\textbf{a}, illustration of states with different filling factors and topological properties. 
\textbf{b}, simulation workflow. A neural network is designed to represent the many-body wavefunction of topological systems, and quantum Monte Carlo is employed to optimize the neural network approaching ground state. 
\textbf{c}, schematic comparison between the neural network approach and the state-of-the-art ED method. The neural network approach circumvents the factorial scaling problem of the Hilbert space that limits the ED method, and consequently achieves better accuracy. }
\end{figure*}

Topological physics has emerged as a thriving branch of modern condensed matter physics, fundamentally altering our understanding of quantum phases of matter. A milestone was the discovery of the quantum Hall effect, where Landau levels in a two-dimensional electron gas under a strong magnetic field give rise to quantized Hall conductance \cite{qhe, fqhe}. This phenomenon unveiled the profound connection between topology and electronic properties, inspiring the theoretical prediction and experimental realization of a diverse range of topological states \cite{haldane_model, z2, ti_rev1, ti_rev2, ti_rev3, ti_rev4}. These materials, distinguished by their topologically nontrivial bulk states and topologically protected edge or surface states, exhibit robust electronic properties that resist disorder and weak perturbations, making them promising candidates for next-generation electronic devices and quantum computing technologies.

Within the realm of integer topological states, integer Chern insulator and $Z_2$ topological insulator in two dimensions stand out as representative examples. The integer Chern insulator in the absence of an external magnetic field, also known as quantum anomalous Hall insulator, breaks time-reversal symmetry and exhibits a quantized Hall conductance  \cite{haldane_model}. Conversely, the $Z_2$ topological insulator preserves time-reversal symmetry, hosting spin-momentum-locked edge states \cite{z2}. Beyond these integer topological insulators, fractional Chern insulator (FCI) emerges as a striking manifestation of topological states in the presence of strong electron-electron interactions \cite{fci_rev1}. FCI can be understood as an analog of the fractional quantum Hall effect but without the need for an external magnetic field. Unlike its integer counterpart, FCI is an inherently many-body phase with excitations obeying fractional statistics, making it of great interest for both fundamental physics and potential applications in topological quantum computing. The recent experimental realization of FCI in moir\'e materials has provided a platform for exploring these exotic states \cite{fci_exp1, fci_exp2, fci_exp3, fci_exp4}.

Despite rapid experimental progress, a comprehensive theoretical understanding of FCI remains elusive. This challenge stems primarily from the strongly correlated nature of FCI, which prevents an adiabatic connection from a FCI state to a single Slater  determinant and necessitates a non-perturbative treatment of the full Hilbert space.  While exact diagonalization (ED) and density matrix renormalization group (DMRG) methods have provided valuable insights \cite{fci_ed1, fci_ed2, fci_ed3, fci_ed4, fci_ed5, fci_ed6}, the computational cost of ED scales exponentially with system size and the DMRG calculation is typically performed within a projected Hilbert space, restricting their application to small systems or limiting achievable accuracy. As a result, there is a compelling need for theoretical approaches to investigate correlated topological phases.

In recent years, deep learning quantum Monte Carlo (QMC) methods have emerged as a promising tool for solving complex electronic structure problems, surpassing the capabilities of many conventional techniques \cite{rbm, ferminet, paulinet}. These approaches utilize neural networks to represent the many-body wavefunctions of quantum systems, enabling unprecedented precision. Neural networks, with their vast number of trainable parameters, allow the representation of intricate wavefunctions that would otherwise be computationally intractable. The combination of neural networks and QMC not only ensures high accuracy but also significantly enhances computational efficiency. This synergy has achieved notable success in diverse systems, including molecules \cite{ferminet, paulinet, lapnet}, periodic systems \cite{wap, ferminet_heg, mess_heg, mess_heg2, deepsolid}, fractional quantum Hall systems in Landau levels \cite{fqhe1, fqhe2} and moir\'e materials \cite{deepsolid_moire, psi_moire}. However, the application of neural network-based QMC methods to topological insulators, particularly those at fractional fillings, remains a significant challenge. The unique interplay of non-trivial topology and strong correlations presents substantial hurdles for neural network representations, leaving the full potential of this approach largely untapped in this critical domain.

In this study, we introduce a deep learning framework for investigating topological systems, focusing on twisted bilayer ${\rm MoTe_2}$ ($t{\rm MoTe_2}$) as a representative example. Moir\'e materials, with their tunable band structures and inherent strong correlations, offer an ideal platform for exploring a diverse range of topological phases. By constructing a neural network architecture tailored to capture the non-trivial topology of the many-body wavefunction, our method accurately identifies and characterizes various topological states, including integer Chern insulators, $Z_2$ topological insulators, and the more challenging FCIs. Notably, our results not only outperform traditional ED methods in both accuracy and efficiency, but also predict phases that align closely with experimental observations. These findings highlight the power of deep learning approaches to provide unbiased solutions to topological quantum matter with strongly correlated many-body effects.

\section{Result}
\begin{figure*}[t!]
\centering
\includegraphics[width=0.95\textwidth]{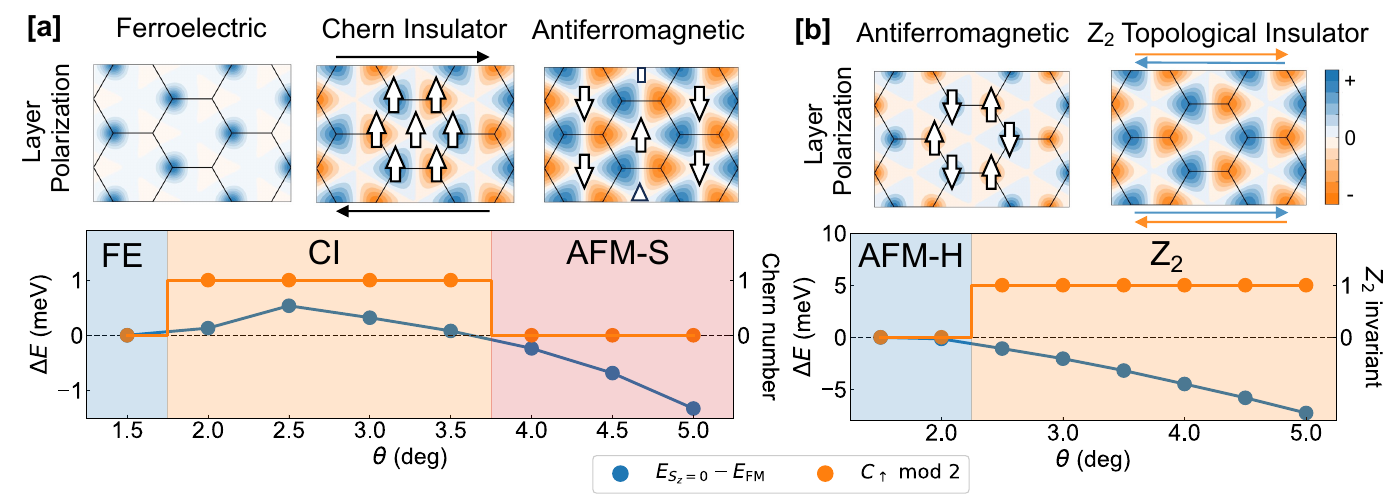}
\caption{\label{fig:ti} \textbf{Integer topological states.} \textbf{a-b}, calculated phase diagram at $n=-1$ and $-2$. $3\times3$ supercells built from rectangular cells and triangular cells are employed for $n=-1$ and $-2$ respectively. Upper panel plots layer polarization density of calculated phases, defined as expectation value of layer-spin projected in $z$ direction. Arrows in the figure denote the physical-spin distribution. Lower panel presents the calculated energy difference $\Delta E=E_{S_z=0}-E_{\rm FM}$ per particle between unpolarized state ${S_z=0}$ and polarized ferromagnetic state, and the spin-up Chern number, $C_\uparrow$, of ground states at different twist angles. FE denotes ferroelectric states in which spin states are nearly degenerated. CI denotes Chern insulator. AFM-S and AFM-H denote the antiferromagnetic phases where physical-spins form stripe pattern at $n=-1$ and honeycomb crystal at $n=-2$, respectively.}
\end{figure*}

\subsection{Deep Learning Simulation}
Fig.~\ref{fig:concept} provides a schematic overview of our deep learning methodology. Accurately solving the many-body Schr\"odinger equation is crucial for investigating the topological properties of quantum systems. However, obtaining a precise solution for strongly correlated systems remains extremely challenging. 
To address this, we extend the DeepSolid neural network architecture \cite{deepsolid}, previously validated for simulating real solids and moir\'e materials, to represent the many-body wavefunctions of topological systems.
As depicted in Fig.~\ref{fig:concept}b, the particle information is fused and passed through a series of deep fully connected neural networks to generate the wavefunction. This architecture effectively captures particle correlations, enabling an accurate representation of the quantum state. Once the wavefunction is constructed, we apply variational Monte Carlo (VMC) to optimize the parameters of the neural network. VMC, with its favorable computational scaling of $\mathcal{O}(N^{3-4})$ and absence of the sign problem, ensures both accuracy and efficiency in our simulations.

Our method offers a versatile and powerful framework for studying topological systems across a wide range of materials. To demonstrate its capabilities, we apply it to ${t\rm MoTe_2}$, a system that has recently attracted significant attention for hosting various topological phases \cite{fci_exp1,fci_exp2, fci_exp3, fci_exp4}. $t{\rm MoTe_2}$ features strong spin-valley locking, leading to a separation of its top valence band from others \cite{mote2_model1, mote2_model2}. The effective single-particle Hamiltonian $H_{\uparrow}$ for spin-up valence electrons  after particle-hole transformation is given by:
\begin{equation}
\begin{gathered}
    H_{\uparrow}(\br)=
    \begin{pmatrix}
    \frac{(-i\nabla-\mathbf{K}_+)^2}{2m} + \Delta_b(\br) & \Delta_T(\br) \\
    \Delta^*_T(\br) & \frac{(-i\nabla-\mathbf{K}_-)^2}{2m}+\Delta_t(\br)
    \end{pmatrix},
\end{gathered}
\label{eq:ham}
\end{equation}  
where $\mathbf{K}_\pm$ denote high-symmetry points in the Brillouin zone. This Hamiltonian has a spinor structure, representing the layer degree of freedom in bilayer systems. The terms $\Delta_{b/t}$ and $\Delta_T$ represent intra-layer and inter-layer potentials, respectively. These potentials form an effective skyrmion field, giving rise to non-trivial band topology. Specific model parameters are taken from Ref.~\cite{fci_ed1}. The full many-body Hamiltonian is expressed as
\begin{equation}
  H_{\rm total} = \sum_iH_{\uparrow}(\br_i) + H_{\downarrow}(\br_i) + \frac{1}{2}\sum_{i\neq j}v_E(\mathbf{r}_i-\mathbf{r}_j),
\end{equation}
where the spin-down Hamiltonian $H_{\downarrow}$ is related to $H_\uparrow$ by time-reversal symmetry. The term $v_E$ represents the Coulomb interaction between electrons in a uniform charge background, handled with Ewald summation. Madelung constant is omitted here.

\begin{figure*}[t!]
\centering
\includegraphics[width=0.95\textwidth]{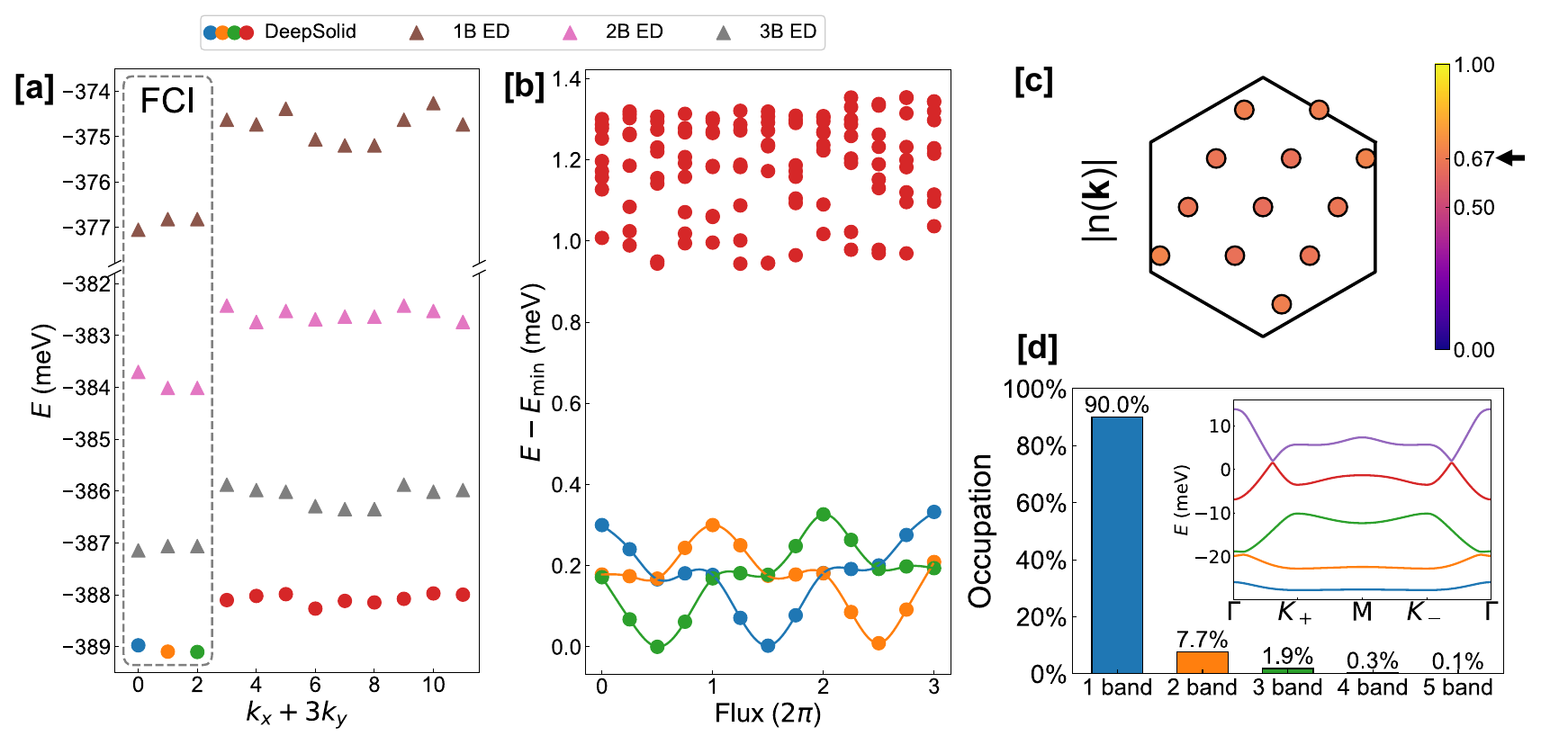}
\caption{\label{fig:fci} \textbf{Fractional Chern insulator.} \textbf{a}, energy spectrum of FCI at $n=-2/3$. ED results are plotted for comparison. Ferromagnetic is assumed and $3\times4$ supercell is used at $\theta=2^\circ$. 
\textbf{b}, energy spectrum for different flux insertions. The three FCI states remain gapped from the higher-lying excited states. As flux is increased, the three FCI states mix with each other.
\textbf{c}, calculated one-body density matrix $\langle n(\bk)\rangle$ at different momentum. Averaging $\langle n(\bk)\rangle$ over three degenerated FCI states reveals nearly uniform distributions. \textbf{d}, band occupation analysis of the neural network wavefunction. Occupation of $i$-th band is defined as $\sum_\bk n_i(\bk)/N_e$.  Inset plots the corresponding single-particle band structures.}
\end{figure*}

\subsection{Integer Topological States}
Topological states are distinguished by nontrivial bulk topological invariants. 
A key quantity in characterizing topological states is the many-body Chern number \cite{mbcn}, defined as the integral of the many-body Berry curvature $\mathbf{F}(\ks)$ over the supercell Brillouin zone (SBZ):
\begin{equation}
\begin{gathered}
C = \frac{1}{2\pi i}\int_{\rm SBZ} \mathbf{F}(\ks)\ {\rm d}\ks,\\
\mathbf{F}(\ks)=\nabla_{\ks}\times\langle \Phi_\ks|\nabla_{\ks}|\Phi_{\ks}\rangle,\\
\end{gathered}
\label{eq:topo}
\end{equation}
where $\Phi_{\ks}$ denotes the supercell-periodic part of the many-body wavefunction at twist momentum $\ks$. In practice, the many-body Chern number can be calculated using rotation symmetry \cite{rotsym} or the single-point formula \cite{single_point1, single_point2} without carrying out the integration explicitly. See method section for details.

To illustrate the capabilities of our neural network method, we first investigate the phase diagram of $t{\rm MoTe_2}$ at integer fillings. In our simulations, we take the $z$-component of the physical-spin to be a good quantum number (i.e., spin U(1) symmetry) and determine the ground state by comparing energies of different spin configurations, specifically the fully spin-polarized and spin-unpolarized states. For filling $n=-1$, as shown in Fig.~\ref{fig:ti}a, we observe the existence of ferroelectric states at small twist angles ($\theta\lesssim1.75^\circ$), where electrons accumulate in one layer, forming spontaneous out-of-plane electric polarization. This behavior arises from the dominance of Coulomb interactions at small twist angles, which drives charge localization. The two spin configurations with, respectively, ferromagnetic and antiferromagnetic (AFM)  order, are nearly degenerate, due to the weak magnetic coupling between localized moments in the moir\'e superlattices with a large period.   As the twist angle increases ($\theta \gtrsim 2.0^\circ$), the spin-polarized ferromagnetic state becomes the definite ground state. Moreover, the system acquires a Chern number of $C=1$, establishing its topological nature. Further increasing the twist angle ($\theta \gtrsim 4.0^\circ$) results in a transition from ferromagnetic to antiferromagnetic state. In contrast to the $120^\circ$ N\'eel order states predicted in Ref.~\cite{mote2_model2} that spontaneously break the spin U(1) symmetry, we only consider spin U(1) symmetric states in our simulation and, therefore, observe the formation of an AFM stripe pattern as shown in Fig.~\ref{fig:ti}a. (See supplementary material for details.)

We further extend our investigation to the $n=-2$ filling, where the system exhibits a distinct behavior compared to the $n=-1$ case. At small twist angles, the system forms an antiferromagnetic honeycomb crystal with trivial topology (see Fig.~\ref{fig:ti}b). As the twist angle increases, the system undergoes a transition to a topologically non-trivial phase. Unlike the ferromagnetic Chern insulator observed at $n=-1$, the system retains time-reversal symmetry and becomes a $Z_2$ insulator around $\theta\simeq 2^\circ$, which is consistent with recent experimental results reporting quantum spin Hall effects in $t{\rm MoTe_2}$ \cite{spin_hall_exp}. These findings underscore the versatility and power of our neural network-based approach, which provides an accurate and efficient tool for discovering topological phases across a range of materials and system settings.

\subsection{Fractional Chern Insulator}
A more challenging scenario is the topological phenomena at fractional fillings, where electron-electron correlation effects are more significant.
In the flat-band limit, where the kinetic energy is quenched, all accessible single-electron configurations represented by Slater determinants become degenerate, necessitating the consideration of the full Hilbert space to accurately capture the system's behavior. This intrinsic degeneracy renders conventional computational approaches, which often rely on perturbative or mean-field approximations, inadequate for studying FCI. 

\begin{figure*}[t!]
\centering
\includegraphics[width=0.95\linewidth]{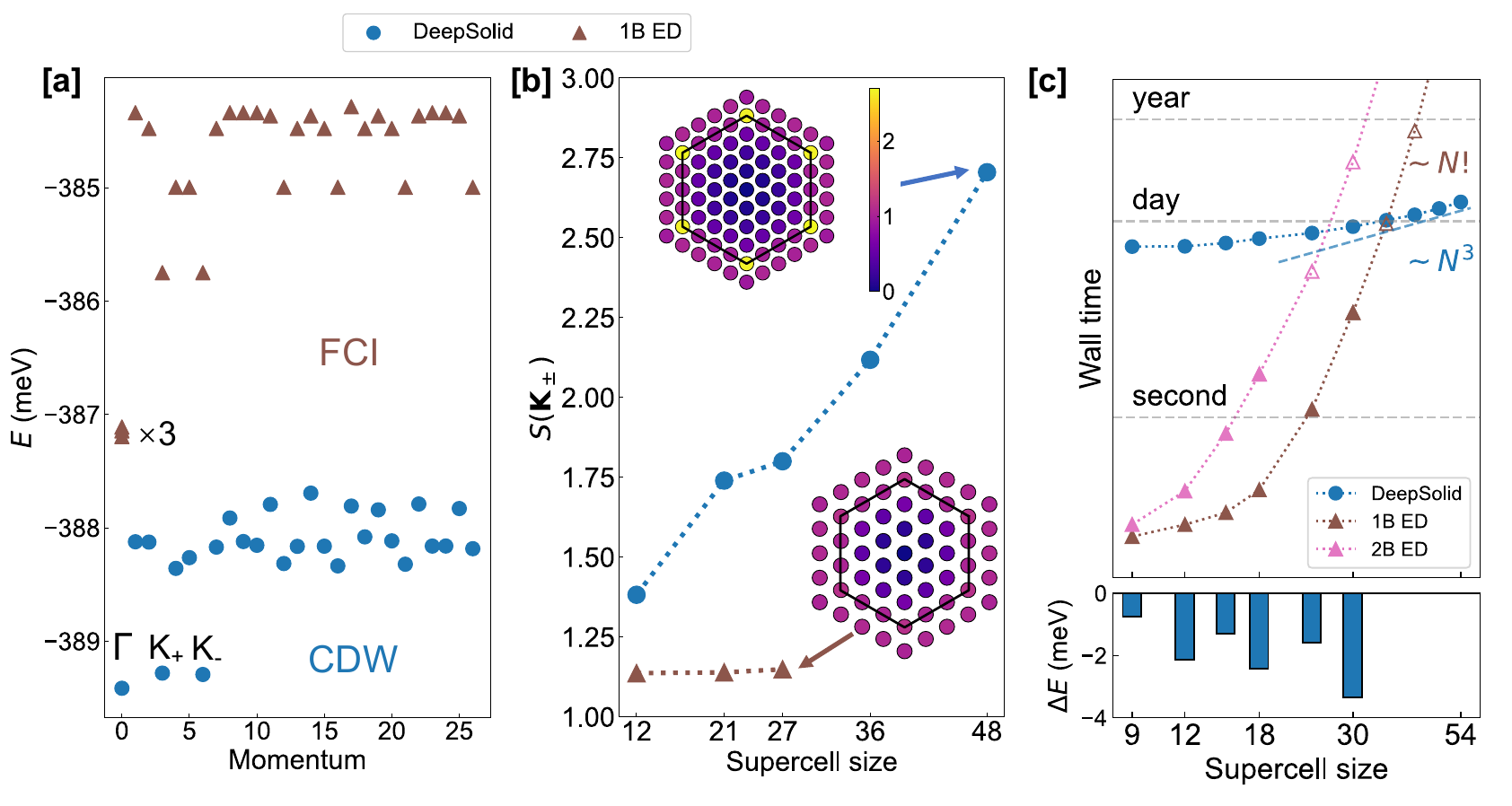}
\caption{\label{fig:fci-cdw} \textbf{FCI-CDW competing and computational cost.} 
\textbf{a}, calculated energy spectrum for $n=-1/3$ filling at $\theta=2^\circ$. $3\sqrt{3}\times3\sqrt{3}$ supercell is employed. DeepSolid finds threefold degenerated ground state with momenta $\Gamma,\mathbf{K}_\pm$, confirming its CDW nature. In contrast, one-band ED find three degenerated state at $\Gamma$, manifesting FCI state instead.
\textbf{b}, heights of the structure factor at $\mathbf{K}_\pm$ points for ground states of different supercell sizes. The insets show the 2D structure factor for a 48-site (DeepSolid) and a 27-site (ED) supercell. The DeepSolid result shows sharp peaks forming at $\mathbf{K}_\pm$ points and growing with supercell size. No significant peaks are found in one-band ED results. Structure factors of ED are taken from~\cite{fci_sf}.
\textbf{c}, comparison of computational cost and energy accuracy of different methods. $n=-1/3$ filling is used as an example. The upper panel presents estimated wall-time: DeepSolid time is based on $10^5$ training steps, while ED time is for a Lanczos eigenvalue calculation (hollow markers indicate estimates). Lower panel plots the energy difference between DeepSolid and one-band ED, $\Delta E = E_\mathrm{DeepSolid} - E_\mathrm{ED}$. 
 }
\end{figure*}

Here we explore the FCI states of $t{\rm MoTe_2}$ at a fractional filling of $n=-2/3$ with our neural network wavefunction approach.
In order to obtain the full energy spectrum, we symmetrize our neural network and construct translation symmetric states with well-defined center-of-mass momentum. (See method section for details.) For simplicity, we assume ferromagnetism and use a $3\times4$ supercell at $\theta=2^\circ$. Our results, presented in Fig.~\ref{fig:fci}, reveal that the ground states exhibit a threefold topological degeneracy, with the degenerate states occupying momentum sectors consistent with the generalized Pauli principle \cite{fci_rev1}, which is a strong indication of FCI states.
To assess the accuracy of our neural network approach, we compare the obtained energy spectra with ED results, as shown in Fig.~\ref{fig:fci}a. Our neural network wavefunction yields energies that significantly outperform even three-band ED calculations, which are already at the computational limit. Furthermore, the threefold degeneracy of the ground states is preserved across various twist boundary conditions, and these degenerate states are permuted during flux insertion with a period of $6\pi$ while remaining well-gapped from higher-energy excitations (Fig.~\ref{fig:fci}b), which confirms the topological nature of the FCI states. The near-uniform density distribution of these states across the Brillouin zone further substantiates their identification as a FCI phase (Fig.~\ref{fig:fci}c). The band occupation analysis in Fig.~\ref{fig:fci}d reveals that the neural network wavefunction encompasses multiple bands, highlighting the method's capability to capture band-mixing effects and achieve high accuracy. We also find similar FCI states at a filling of $n=-3/5$ and the calculated results are provided in supplementary material.

Moving forward, we delve into the delicate competition between charge density wave (CDW) and FCI phases, as both phases can exhibit similar ground state degeneracies but differ in their topological properties, making their distinction a significant challenge \cite{fci_rev1}. 
In analogy to the fractional quantum Hall effect, one might expect FCI states to emerge at numerous fractional fillings, particularly in the $t$MoTe$_2$ system with a nearly flat topological band, where its wavefunction can resemble that of the generalized Landau level \cite{li2025variation}. However, experiments in $t{\rm MoTe_2}$ have revealed a more complex picture. Only a limited number of FCI states, such as those at $n=-2/3$ and $-3/5$, have been observed, with other anticipated FCI phases instead manifesting as CDW states \cite{fci_exp1, fci_exp2, fci_exp3}. Notably, at $n=-1/3$, ED studies suggested an FCI state near the magic angle $\theta \approx 2^\circ$ \cite{fci_ed1, fci_ed5}, which has not been observed in experiments \cite{fci_exp6}.

To resolve the competition between CDW and FCI phases, we perform large-scale simulations for $n=-1/3$ near the magic angle, exceeding previous studies in system size and fully incorporating band-mixing effects as our calculation does not rely on band projection. The neural-network-computed spectrum, alongside one-band ED result, is presented in Fig.~\ref{fig:fci-cdw}a. Notably, the neural network result reveals a threefold degeneracy at both the $\Gamma$ and $\bK_{\pm}$ high-symmetry points, strongly indicative of a CDW ground state. In contrast, the ED method predicts degeneracy at $\Gamma$, consistent with FCI character. Further corroborating the CDW order, the structure factor from neural network calculations (Fig.~\ref{fig:fci-cdw}b) exhibits sharp crystalline signatures, with pronounced Bragg peaks emerging at $\mathbf{K}_{\pm}$ and heights growing linearly with respect to system size. This crystalline behavior persists at $\theta = 2^\circ$ near the magic angle, deviating sharply from the one-band approximation and underscoring the critical role of band-mixing effects even in the flat-band regime. Our findings, which contradict ED calculations and previous works \cite{fci_ed1, fci_ed5} but align closely with experimental observations \cite{fci_exp6}, highlight the capability of our method to model the untruncated Hilbert space with minimal finite-size errors.

To systematically address the computational demands of large-scale systems, we benchmark the scaling behavior and accuracy of our approach in Fig.~\ref{fig:fci-cdw}c. Our method exhibits an asymptotic scaling of $\mathcal{O}(N^3)$ with respect to system size, in stark contrast to the factorial scaling of ED. While communication overhead dominates in our method for small systems, resulting in a near-constant cost, its cubic scaling for larger systems represents a dramatic improvement. Moreover, our method consistently achieves lower variational energies across all simulated system sizes to which ED is applicable, indicating higher accuracy. This superior scaling and accuracy allow us to overcome the limitations of ED and explore system sizes far beyond its reach, which is crucial for understanding the intricate physics of strongly correlated topological phases that were otherwise intractable with traditional methods.

\section{Conclusions and Outlook}
In this work, we introduced a deep-learning-based QMC approach for simulating integer and fractional topological states, leveraging neural networks to represent many-body wavefunctions. By integrating the expressiveness of neural networks with the accuracy and efficiency of QMC, our method successfully captures various topological phases in $t{\rm MoTe_2}$ at both integer and fractional fillings. Our approach surpasses conventional exact diagonalization in both accuracy and scalability, offering an efficient framework to investigate strongly correlated topological systems beyond current computational limits.

While our method offers a powerful tool for topological studies, we primarily focus on ground-state properties. A natural extension for this approach is to investigate excited states \cite{excite}, as well as dynamical properties \cite{time_evolution} and finite-temperature effects \cite{finite_temperature}. These advancements would enable a comprehensive exploration of the full energy spectrum and deepen our understanding of exotic states in correlated topological matter.

Looking forward, we see many fascinating research avenues to be explored. Notably, the investigation of fractional excitations \cite{composite} and anyon physics  holds particular interest. Unlike conventional fermions and bosons, anyons obey fractional exchange statistics, where their wavefunction acquires a nontrivial phase factor upon braiding. 
Recent experiments have reported evidence of fractional quantum spin Hall insulator with possible non-abelian anyons in $t$MoTe$_2$ \cite{spin_hall_exp}, yet their theoretical underpinnings remain an open question. 
Another compelling direction is the exploration of superconductivity in moir\'e materials. While superconducting states have recently been observed in similar $t{\rm WSe_2}$ systems \cite{wse2_sc, wse2_sc2}, their underlying pairing mechanisms remain mysteries. 
Addressing these questions requires a computational approach that is both highly accurate and efficient. Our method stands as a strong candidate for this challenge, providing a robust foundation for exploring complex quantum phases in strongly correlated systems and providing more insights into topological physics.

\section{Method}
\subsection{Moir\'e Hamiltonian}
Moir\'e systems contain numerous electrons, making \textit{ab-initio} calculations impractical. An effective continuum model is usually derived to simplify the problem, which only considers the most active low-energy states. For $t{\rm MoTe_2}$ system, a two-component Hamiltonian is constructed for each spin channel, accounting for the orbitals in the top and bottom layers respectively. And the periodic potential term $\Delta$ reads
\begin{equation}
\begin{gathered}
    \Delta_{b/t}(\br)=-2V\sum_{i=1,3,5}\cos(\mathbf{g}_i\cdot\br\pm\delta), \\
    \Delta_{T}(\br)=\omega(1+e^{i\mathbf{g}_2\cdot\br}+e^{i\mathbf{g}_3\cdot\br}),
\end{gathered}
\label{eq:mag}
\end{equation}
where $\mathbf{g}_i=\frac{4\pi}{\sqrt{3}a_M}(\cos(\frac{\pi(i-1)}{3}),\sin(\frac{\pi(i-1)}{3}))$ are the primitive cell reciprocal lattice vectors, and $V,\omega,\delta$ are parameters fitted from density functional results \cite{fci_ed1}.

\subsection{Continuous Spin Neural Network}
The two-component Hamiltonian for $t{\rm MoTe_2}$ does not commute with the $z$-component of the layer-spin, implying that the layer-spin of each particle must be treated as a dynamic variable. In QMC simulations, this introduces a critical bottleneck: discrete spin variable $s_z$ can reduce sampling efficiency due to high rejection rates. 
To address this, the continuous spin technique is employed  which maps the discrete spins $s_z$ onto a continuous variable $s\in [0,2\pi)$ \cite{continuous_spin}. The single-particle wavefunction is reparametrized as:
\begin{equation}
\begin{gathered}
    \phi(\br,s) = \phi^b(\br)e^{is}+\phi^t(\br)e^{-is},
\end{gathered}
\label{eq:continuous_spin}
\end{equation}
where $\phi^{b/t}$ denote corresponding spatial orbitals. By unifying spin and spatial coordinates into a continuous framework, this approach achieves highly efficient sampling of both variables. To incorporate electron correlations, these single-particle orbitals $\phi^{b/t}(\br_i)$ are further promoted to neural network orbitals $\phi^{b/t}(\br_i;\br_{\neq i})$. See supplementary material for a detailed network structure.

\subsection{Many-Body Topological Invariants \label{sec:mbti}}
The many-body Chern number is a fundamental topological invariant for correlated insulators, encoding quantized Hall conductance and robust edge states \cite{mbcn}. Conventionally, it is computed by integrating the Berry curvature over the supercell Brillouin zone, which is usually impractical. To overcome this, we employ the single-point formula which leverages polarization operators to extract Chern number \cite{single_point1, single_point2}. For spin channel $\alpha\ (\alpha=\uparrow,\downarrow)$, $\rho_\alpha$ is defined as:
\begin{equation}
\begin{gathered}
    \rho_\alpha=\frac{\langle\Psi|Z_{\alpha}(\mathbf{G}_1+\mathbf{G}_2)|\Psi\rangle}{\langle\Psi|Z_{\alpha}(\mathbf{G}_1)|\Psi\rangle\langle\Psi|Z_{\alpha}(\mathbf{G}_2)|\Psi\rangle}, \\
    Z_\alpha(\mathbf{G})=e^{i\mathbf{G}\cdot\sum_i\br^\alpha_i}\ ,\\
\end{gathered}
\label{eq:single}
\end{equation}
where $\mathbf{G}$ denotes supercell reciprocal lattice vectors. It can be proved that the phase angle of $\rho_\alpha$ reduces to Chern number at the thermodynamic limit (TDL), which reads
\begin{equation}
\begin{gathered}
    {\rm Arg}[\rho_\alpha]/\pi\xRightarrow[\text{TDL}]{} C_{\alpha} \ {\rm mod} \ 2\ .
\end{gathered}
\label{eq:single_tdl}
\end{equation}
Eq.~\eqref{eq:single} only requires a single many-body wavefunction which efficiently determines Chern number up to modulus 2. Furthermore, it is crucial to consider the role of time-reversal symmetry when characterizing the system's topological properties. In systems with broken time-reversal symmetry, a non-zero Chern number indicates a Chern insulator. In contrast, for time-reversal symmetric systems, the Chern numbers vanish, and the many-body $Z_2$ invariant $\nu=(C_\uparrow-C_\downarrow)/2$ becomes the relevant topological index.

\subsection{Translational Symmetric State}
Periodic systems have two types of translational symmetries, which read
\begin{equation}
\begin{gathered}
    \Psi(\br_1+\bl,...,\br_N+\bl)=e^{{i\kp\cdot\bl}}\Psi(\br_1,...,\br_N)\ , \\
    \Psi(\br_1+\bL,...,\br_N)=e^{{i\ks\cdot\bL}}\Psi(\br_1,...,\br_N)\ .
\end{gathered}
\label{eq:periodic_con}
\end{equation}
Here, $\ks$ denotes the twist momentum associated with a translation of any electron by a supercell lattice vector $\bL$, and $\kp$ denotes the center-of-mass momentum corresponding to a simultaneous translation of all electrons by a primitive cell lattice vector $\bl$.

To identify quantum states with different symmetries, we construct $\bk_p$-symmetric neural network wavefunctions, which take the form:
\begin{equation}
\begin{gathered}
\Psi_{\kp}(\br_1,...,\br_N)=\sum_{\bl\in {\rm supercell}}e^{i\kp\cdot\bl}\Psi_{\rm Net}(\br_1-\bl,...,\br_N-\bl).
\end{gathered}
\label{eq:kp_symmetric}
\end{equation}
where the summation is constrained in a single supercell, and $\Psi_{\rm Net}$ only possesses supercell translational symmetry. And twist momentum $\bk_s$ can be simply fixed by multiplying an overall phase factor $e^{i\bk_s\cdot\sum_i\br_i}$. This construction provides the most general $\bk_p$ symmetric wavefunctions, but it requires multiple forward calculations of the neural network, which increases computation costs.

\subsection{Workflow and Computational Details}

In our simulations, the neural network is randomly initialized, with spatial positions and continuous spins uniformly distributed at the outset. We then apply the variational quantum Monte Carlo method, which samples configurations from the neural network wavefunction, computes energy gradients, and optimizes the network. Specifically, we employ the Kronecker-factored approximate curvature optimizer \cite{kfac}. To assess the robustness of our method, we explored different simulation parameters and conducted multiple runs, which show no notable differences. To accelerate the simulations, we utilize the Forward Laplacian \cite{lapnet} and fast-update technique \cite{fast_update}. Most simulations were performed on eight H20 GPUs, typically completing within a few hours. Orbital analysis follows the approach outlined in Ref.~\cite{pyqmc}. The calculated energy and topological results are provided in the supplementary material.

\section*{Acknowledgements}
We want to thank Liang Fu, Shiwei Zhang, Long Ju, Jie Shan, Yang Xu, Haruki Watanabe and Jiaqi Cai for helpful discussions. We want to thank Aidan P. Reddy and Timothy Zaklama for sharing ED data. We want to thank Yubo Yang and Agnes Valenti for early discussions. We want to thank ByteDance Research Group for inspiration and encouragement. This work is directed and supported by Hang Li and ByteDance Research. J.C. is supported by the National Key R\&D Program of China under Grant No.
2021YFA1400500 and the National Natural Science Foundation of China under Grant No. 12334003.
F. W. is supported by National Key Research and Development Program of China (Grants  No. 2022YFA1402401 and No. 2021YFA1401300), and National Natural Science Foundation of China (Grant No. 12274333). The ED calculations have been performed on the supercomputing system in the Supercomputing Center of Wuhan University.

\bibliography{reference}

\end{document}



\title{Supporting Information: Deep Learning Sheds Light on Integer and Fractional Topological Insulators}
\author{Xiang Li}
\email{lixiang.62770689@bytedance.com}
\affiliation{ByteDance Research}
\author{Yixiao Chen}
\affiliation{ByteDance Research}
\author{Bohao Li}
\affiliation{School of Physics and Technology,  Wuhan University}

\author{Haoxiang Chen}
\affiliation{ByteDance Research}
\affiliation{School of Physics, Peking University}

\author{Fengcheng Wu}
\email{wufcheng@whu.edu.cn}
\affiliation{School of Physics and Technology, Wuhan University}

\author{Ji Chen}
\email{ji.chen@pku.edu.cn}
\affiliation{School of Physics, Peking University}

\author{Weiluo Ren}
\email{renweiluo@bytedance.com}
\affiliation{ByteDance Research}

\maketitle

\onecolumngrid

\section{Moir\'e Hamiltonian}
Continuum model of $t{\rm MoTe_2}$ reads
\begin{equation}
\begin{gathered}
    H_{\uparrow}(\br)=
    \begin{pmatrix}
    \frac{(-i\nabla-\mathbf{K}_+)^2}{2m} + \Delta_b(\br) & \Delta_T(\br) \\
    \Delta^*_T(\br) & \frac{(-i\nabla-\mathbf{K}_-)^2}{2m}+\Delta_t(\br)
    \end{pmatrix},\\
        H_{\downarrow}(\br)=
    \begin{pmatrix}
    \frac{(-i\nabla+\mathbf{K}_+)^2}{2m} + \Delta_b(\br) & \Delta_T^*(\br) \\
    \Delta_T(\br) & \frac{(-i\nabla+\mathbf{K}_-)^2}{2m}+\Delta_t(\br)
    \end{pmatrix},\\
        \Delta_{b/t}(\br)=-2V\sum_{i=1,3,5}\cos(\mathbf{g}_i\cdot\br\pm\delta), \\
    \Delta_{T}(\br)=\omega(1+e^{i\mathbf{g}_2\cdot\br}+e^{i\mathbf{g}_3\cdot\br}),\\
    \mathbf{g}_i=\frac{4\pi}{\sqrt{3}a_M}(\cos(\frac{\pi(i-1)}{3}),\sin(\frac{\pi(i-1)}{3})),\  \mathbf{K}_+=\frac{\bg_1+\bg_2}{3},\ \mathbf{K}_-=\frac{\bg_1+\bg_6}{3},\\
        H_{\rm total} = \sum_iH_{\uparrow}(\br_i) + H_{\downarrow}(\br_i) + \frac{1}{2}\sum_{i\neq j}v_E(\br_i-\br_j)+\frac{N}{2}v_M,
\end{gathered}
\label{eq:ham}
\end{equation}
where $a_M=\frac{a_0}{2\sin(\theta/2)}$ denotes moir\'e length at twist angle $\theta$. $v_E$ represents Coulomb interaction between different electrons in a uniform charge background, and reads \cite{2dewald}
\begin{equation}
\begin{gathered}
    v_E(\br)=\sum_{\mathbf{L}}\frac{{\rm erfc}(\sqrt{\gamma}|\br-\mathbf{L}|}{\epsilon|\br-\mathbf{L}|} + \frac{2\pi}{\epsilon A}\sum_{\mathbf{G}\neq 0}\frac{\exp(i\mathbf{G}\cdot\br)}{|\mathbf{G}|}{\rm erfc}(\frac{|\mathbf{G}|}{2\sqrt{\gamma}})-\frac{2}{\epsilon A}\sqrt{\frac{\pi}{\gamma}},
\end{gathered}
\label{eq:ve}
\end{equation}
where $\gamma$ denotes parameter used in Ewald summations, and $A$ is the supercell area. Madelung constant $v_M$ represents image Coulomb interactions and is defined as 
\begin{equation}
\begin{gathered}
   v_M=\lim_{\br\rightarrow0} \left[v_E({\br})-\frac{1}{\epsilon|\br|}\right]=\sum_{\mathbf{L}\neq0}\frac{{\rm erfc}(\sqrt{\gamma}|\mathbf{L}|)}{\epsilon|\mathbf{L}|} - \frac{2}{\epsilon}\sqrt{\frac{\gamma}{\pi}}+\frac{2\pi}{\epsilon A}\sum_{\mathbf{G}\neq0}\frac{{\rm erfc}(\frac{|\mathbf{G}|}{2\sqrt{\gamma}})}{|\mathbf{G}|} -\frac{2}{\epsilon A}\sqrt{\frac{\pi}{\gamma}}\ .
\end{gathered}
\label{eq:mad}
\end{equation}

Specific parameters are listed in Tab.~\ref{tab:Model Parameters} and taken from Ref.~\cite{fci_ed1}.
\begin{table}[H]
  \centering
\begin{tabularx}{0.6\textwidth}{Y|Y|Y|Y|Y|Y}
\hline
\hline
 $a_0\ ({\rm nm})$ &$m\ (m_e)$ & $V$ (meV) & $\omega$ (meV) & $\delta$ & $\epsilon$ \\
      \hline
 0.352 & 0.62 & 11.2 & -13.3 & $-91^\circ$ & 10\\
\hline
\hline
\end{tabularx}
\caption{Parameters of $t{\rm MoTe_2}$.}
\label{tab:Model Parameters}
\end{table}

\section{Hyperparameters}
Hyperparameters used in simulations are listed in Tab.~\ref{tab:ge_hyper}, including neural network dimensions and quantum Monte Carlo settings. Compared to previous parameter settings, we use a larger gradient clipping window and increased damping in optimization to enhance the performance of the simulation. Different choices of other hyperparameters are tested and have limited effects on the results. 

Notably, a single determinant is employed in our network to represent the many-body wavefunction, in contrast to the widely used multi-determinant forms. It is well known that fractional quantum Hall states exhibit strong correlations, necessitating a full treatment of the Hilbert space, which consists of numerous determinants. To prove that the neural network wavefunction provides a compact representation of quantum Hall states and scales efficiently to larger systems, we adopt a single-determinant formulation. While accumulating more determinants can enhance accuracy, this approach may quickly become impractical as the system size increases.
\begin{table}[H]
  \centering
\begin{tabularx}{0.9\textwidth}{C{6cm}|Y||C{6cm}|Y}
\hline
\hline
Hyperparameter & Value & Hyperparameter & Value\\
\hline
Dimension of one-electron layer & 256 & Dimension of two-electron layer & 32 \\
Number of layers  & 4 & Number of determinants & 1\\
Optimizer & KFAC & Learning rate & 3e-3\\
Learning rate decay & 1 & Learning rate delay & 1e4 \\
Damping & 3e-4 & Constrained norm of gradient & 1e-3 \\
Momentum of optimizer & 0.0 & Batch size & 4096 \\
Number of training steps & 5e4 & Clipping window of gradient & 20 \\
MCMC burn in & 1e3 & MCMC steps between each iteration & 20 \\
MCMC move width & 2e-2 & Target MCMC acceptance & 55\% \\
Precision & Float32 & Number of inference steps & 5e3 \\ 
\hline
\hline
\end{tabularx}
\caption{Recommended hyperparameters.
}
\label{tab:ge_hyper}
\end{table}

\section{Neural network wavefunction}
DeepSolid is employed to construct many-body wavefunction \cite{deepsolid}. Given the spatial positions and spin variables of electrons, we first transform all spatial information to be periodic \cite{ferminet_heg}. This includes the single-particle distance 
$\mathbf{h}_e$ and the relative distance between two particles $\mathbf{h}_{ee'}$, which are given by:
\begin{equation}
\begin{gathered}
    \mathbf{h}=(\br,|\br|)\rightarrow (\sum_i\bl_i\sin(\br\cdot\bg_i), \sum_i\bl_i\cos(\br\cdot\bg_i),d(\br)),\\
    4\pi^2 d^2(\br)=\sum_{ij}\sin(\br\cdot\bg_i)\sin(\br\cdot\bg_j)\bl_i\cdot\bl_j+[1-\cos(\br\cdot\bg_i)][1-\cos(\br\cdot\bg_j)]\bl_i\cdot\bl_j\ .
\end{gathered}
\label{eq:periodic_dis}
\end{equation}
Here $\bl,\bg$ denote lattice vectors and reciprocal lattice vectors, respectively. The periodic distance features of different electrons are then concatenated to form a collective feature vector, which reads
\begin{equation}
\begin{gathered}
    \mathbf{f}_e^\alpha={\rm concat}(\mathbf{h}^\alpha_e,\bg^\uparrow,\bg^\downarrow,\bg_e^{\alpha,\uparrow},\bg_e^{\alpha,\downarrow}), \\
    (\bg^\uparrow,\bg^\downarrow)=(\sum_e\mathbf{h}_e^\uparrow,\sum_e\mathbf{h}_e^\downarrow),\\
    (\bg_e^{\alpha,\uparrow},\bg_e^{\alpha\downarrow})=(\sum_{e'}\mathbf{h}_{ee'}^{\alpha\uparrow},\sum_{e'}\mathbf{h}_{ee'}^{\alpha\downarrow}),
\end{gathered}
\label{eq:cat_feature}
\end{equation}
where $\alpha$ denotes spin index $(\uparrow, \downarrow)$. These collective features $\mathbf{f}$ are then passed through a series of fully-connected layers to output the quasi-orbitals $\phi_i(\br_j;\br_{\neq j})$. Since layer-spin variables are introduced in our study, we double the neural network output to express the spatial orbitals of opposite spinors, and the orbitals are expressed as:
\begin{equation}
\begin{gathered}
\phi^b(\br_i;\br_{\neq i})e^{is_i}+\phi^t(\br_i;\br_{\neq i})e^{-is_i},
\end{gathered}
\label{eq:continuous_spin}
\end{equation}
where $\br_{\neq i}$ denotes all the electron positions except $\br_i$.  

For integer fillings, we also introduce an envelope function  $e^{i\bk\cdot\br}$ to the orbitals which can improve optimization, and all selected $\bk$ points form a closed shell in Brillouin zone. 

For fractional fillings, we include all possible $\bk$ points within first Brillouin zone into considerations and the envelope functions read $\sum_j\pi_{ij}e^{i\bk_j\cdot\br}$ where $\pi$ are trainable parameters. 

Finally, these orbitals are combined with a Slater determinant to form the final wavefunction,
\begin{equation}
\begin{gathered}
    \Psi=\det[\phi_i^\uparrow(\br_j^\uparrow,s_j^\uparrow;\br_{\neq j},s_{\neq j})]\det[\phi_i^\downarrow(\br_j^\downarrow,s_j^\downarrow;\br_{\neq j},s_{\neq j})],\\
    \phi_i(\br_j,s_j;\br_{\neq j}, s_{\neq j})=[\phi^b_i(\br_j,s_j;\br_{\neq j},s_{\neq j})e^{is_j}+
    \phi^t_i(\br_j,s_j;\br_{\neq j},s_{\neq j})e^{-is_j}]\ {\rm envelope}_i(\br_j).
\end{gathered}
\label{eq:final_wf}
\end{equation}

\section{Expectation Formula}
The $t{\rm MoTe_2}$ Hamiltonian contains layer-spin degree of freedom. Simultaneously handling continuous spatial positions and discrete spin variable can cause serious sampling problems in Monte Carlo. As a solution, we use continuous spin technique to express layer-spin, and neural network is employed to express spatial part of each spinor. 

The neural network wavefunction reads
\begin{equation}
\begin{gathered}
\label{eq:net_arc}
    \Psi_{\rm Net}(\mathbf{R}, S)={\rm det}\left|
    \begin{matrix}
    \phi_1^b(\br_1;\br_{\neq1})e^{is_1}+\phi_1^t(\br_1;\br_{\neq1})e^{-is_1}  & \cdots & \phi_N^b(\br_1;\br_{\neq1})e^{is_1}+\phi_N^t(\br_1;\br_{\neq1})e^{-is_1}  \\
    \cdot & & \cdot\\
    \cdot & & \cdot\\
    \cdot & & \cdot\\
    \phi_1^b(\br_N;\br_{\neq N})e^{is_N}+\phi_1^t(\br_N;\br_{\neq N})e^{-is_N}  & \cdots & \phi_N^b(\br_N;\br_{\neq N})e^{is_N}+\phi_N^t(\br_N;\br_{\neq N})e^{-is_N}  \\
    \end{matrix}
    \right|\ .
\end{gathered}
\end{equation}
$t{\rm MoTe_2}$ Hamiltonian contains many terms coupled to layer-spin, and the expectation formula of layer-spin $\sum_i\sigma_i^a$  ($a=x,y,z$) reads
\begin{equation}
\begin{gathered}
\label{eq:layer_spin}
    \frac{\langle\Psi|\sum_i\sigma^a_i|\Psi\rangle}{\langle\Psi|\Psi\rangle}=\int d\bx_1d\bx_2...d\bx_N \sum_i|\Psi|^2\frac{\Psi^a_i(\bx_1,...,\bx_N)}{\Psi(\bx_1,...,\bx_N)},
\end{gathered}
\end{equation}
where $\bx$ denotes position $\br$ and layer-spin $s$ together. $\Psi_i^a$ denotes the original wavefunction except its $i$-th row is modified by layer-spin operator $\sigma^a$. The specific formula reads
\begin{equation}
\begin{gathered}
\label{eq:}
    \Psi ^x_1(\mathbf{R}, S)={\rm det}\left|
    \begin{matrix}
    \phi_1^t(\br_1;\br_{\neq1})e^{is_1}+\phi_1^b(\br_1;\br_{\neq1})e^{-is_1}  & \cdots & \phi_N^t(\br_1;\br_{\neq1})e^{is_1}+\phi_N^b(\br_1;\br_{\neq1})e^{-is_1}  \\
    \cdot & & \cdot\\
    \cdot & & \cdot\\
    \cdot & & \cdot\\
    \phi_1^b(\br_N;\br_{\neq N})e^{is_N}+\phi_1^t(\br_N;\br_{\neq N})e^{-is_N}  & \cdots & \phi_N^b(\br_N;\br_{\neq N})e^{is_N}+\phi_N^t(\br_N;\br_{\neq N})e^{-is_N}  \\
    \end{matrix}
    \right|\ ,\\
        \Psi ^y_1(\mathbf{R}, S)={\rm det}\left|
    \begin{matrix}
    -i\phi_1^t(\br_1;\br_{\neq1})e^{is_1}+i\phi_1^b(\br_1;\br_{\neq1})e^{-is_1}  & \cdots & -i\phi_N^t(\br_1;\br_{\neq1})e^{is_1}+i\phi_N^b(\br_1;\br_{\neq1})e^{-is_1}  \\
    \cdot & & \cdot\\
    \cdot & & \cdot\\
    \cdot & & \cdot\\
    \phi_1^b(\br_N;\br_{\neq N})e^{is_N}+\phi_1^t(\br_N;\br_{\neq N})e^{-is_N}  & \cdots & \phi_N^b(\br_N;\br_{\neq N})e^{is_N}+\phi_N^t(\br_N;\br_{\neq N})e^{-is_N}  \\
    \end{matrix}
    \right|\ ,\\
        \Psi ^z_1(\mathbf{R}, S)={\rm det}\left|
    \begin{matrix}
    \phi_1^b(\br_1;\br_{\neq1})e^{is_1}-\phi_1^t(\br_1;\br_{\neq1})e^{-is_1}  & \cdots & \phi_N^b(\br_1;\br_{\neq1})e^{is_1}-\phi_N^t(\br_1;\br_{\neq1})e^{-is_1}  \\
    \cdot & & \cdot\\
    \cdot & & \cdot\\
    \cdot & & \cdot\\
    \phi_1^b(\br_N;\br_{\neq N})e^{is_N}+\phi_1^t(\br_N;\br_{\neq N})e^{-is_N}  & \cdots & \phi_N^b(\br_N;\br_{\neq N})e^{is_N}+\phi_N^t(\br_N;\br_{\neq N})e^{-is_N}  \\
    \end{matrix}
    \right|\ .
\end{gathered}
\end{equation}
Since only one row of determinant is modified, fast-update method can be used to accelerate the calculation \cite{fast_update}.

For spin-orbital coupling term $\sum_{i,a}\sigma^a_i\cdot\nabla_i^a$, the formula reads
\begin{equation}
\begin{gathered}
\label{eq:soc}
    \frac{\langle\Psi|\sum_{i,a}\sigma^a_i\cdot\nabla_i^a|\Psi\rangle}{\langle\Psi|\Psi\rangle}=\int d\bx_1d\bx_2...d\bx_N \sum_{i,a}|\Psi|^2\frac{\nabla_i^a\Psi^a_i(\bx_1,...,\bx_N)}{\Psi(\bx_1,...,\bx_N)}.
\end{gathered}
\end{equation}

\section{Integer Filling Result}
For $n=-1$, we choose an enlarged rectangular cell as the primitive cell, which contains two particles. For $n=-2$, we choose triangular cell as the primitive cell, which contains two particles. A $3\times 3$ supercell is constructed from the selected primitive cell in both cases. The specific geometry is given in Tab.~\ref{tab:int_geo}
\begin{table}[H]
  \centering
\begin{tabularx}{0.6\textwidth}{Y|C{3cm}|C{3cm}}
  \hline
  \hline
 & $n=-1$ &   $n=-2$  \\
  \hline
  Lattice vector & $\mathbf{l}_1=(\sqrt{3}, 0)$  & $\mathbf{l}_1=(\sqrt{3}/2, -1/2)$ \\
  & $\mathbf{l}_2=(0, 1)$  & $\mathbf{l}_2=(0, 1)$ \\
  \hline
  \hline
\end{tabularx}
\caption{Geometry of Integer Fillings. The lattice vectors are scaled by corresponding moir\'e length $a_M$.}
\label{tab:int_geo}
\end{table}

Physical-spins are fixed in our simulation, forming two possible spin states $S_z=0,1$ per primitive cell. Calculated results at each twist angle are given in Tab.~\ref{tab:int_en}.
\begin{table}[H]
  \centering
\begin{tabularx}{0.95\textwidth}{C{2cm}|Y|Y|Y|Y|Y|Y|Y|Y}
  \hline
  \hline
  \multicolumn{9}{c}{$n=-1$} \\
  \hline
  $\theta$ (deg) & 1.5 & 2.0 & 2.5 & 3.0 & 3.5& 4.0 & 4.5& 5.0\\
  \hline
  $E_{S_z=0}$ (meV) & -52.9592(1)& -52.6066(1) & -53.4720(1) & -54.9176(2) & -56.1268(2) & -57.1045(3) & -57.9037(4) & -58.5860(4)\\
  $E_{S_z=1}$ (meV) &-52.9593(1) & -52.7362(1) & -54.0077(1) & -55.2370(1)& -56.2056(2) & -56.8668(3) & -57.218(2) & -57.2570(4)\\
  \hline
    \multicolumn{9}{c}{$n=-2$} \\
  \hline
  $\theta$ (deg) & 1.5 & 2.0 & 2.5 & 3.0 & 3.5& 4.0 & 4.5& 5.0\\
  \hline
  $E_{S_z=0}$ (meV) & -59.4784(2)& -60.4936(3) & -62.5960(5) & -64.577(1) & -66.2478(6) & -67.5809(9) & -68.528(1) & -69.196(1)\\
  $E_{S_z=1}$ (meV) &-59.4663(2) & -60.3398(2) & -61.5097(3) & -62.521(1)& -63.0513(9) & -63.0938(9) & -62.716(1) & -61.922(1)\\
  
  \hline
  \hline
\end{tabularx}
\caption{Calculated energy per particle at integer fillings. }
\label{tab:int_en}
\end{table}

Single-point formula is employed to calculate Chern number, which defines $\rho_\alpha$ as 
\begin{equation}
\begin{gathered}
    \rho_\alpha=\frac{\langle\Psi|Z_{\alpha}(\mathbf{G}_1+\mathbf{G}_2)|\Psi\rangle}{\langle\Psi|Z_{\alpha}(\mathbf{G}_1)|\Psi\rangle\langle\Psi|Z_{\alpha}(\mathbf{G}_2)|\Psi\rangle}, \\
    Z_\alpha(\mathbf{G})=e^{i\mathbf{G}\cdot\sum_i\br^\alpha_i}\ .\\
\end{gathered}
\label{eq:single}
\end{equation}

Phase angles of $\rho_\alpha$ equal Chern numbers up to modulus 2, and are given in Tab.~\ref{tab:int_chern}.
\begin{table}[H]
  \centering
\begin{tabularx}{0.95\textwidth}{C{3cm}|Y|Y|Y|Y|Y|Y|Y|Y}
  \hline
  \hline
  \multicolumn{9}{c}{$n=-1$} \\
  \hline
  $\theta$ (deg) & 1.5 & 2.0 & 2.5 & 3.0 & 3.5& 4.0 & 4.5& 5.0\\
  \hline
    $S_z=0,$ ${\rm\ Arg(\rho_\uparrow)\ /\ \pi}$ & 0.000 & 0.020 & 0.003 & 0.001 & 0.000 & -0.001 & -0.001 & 0.001\\
    $S_z=1,$${\rm \ Arg(\rho_\uparrow)\ /\ \pi}$ & -0.008 & 0.997 & 1.000 & 0.999 & 0.998 & -0.991 & 0.990 & -0.990\\
    \hline
      \multicolumn{9}{c}{$n=-2$} \\
    \hline
    $S_z=0,$ ${\rm \ Arg(\rho_\uparrow)\ /\ \pi}$ & 0.000 & -0.002 & 0.986 & -0.985 & -0.959 & 0.990 & 0.944 & 0.968\\
  \hline
  \hline
\end{tabularx}
\caption{Calculated phase angle of $\rho_\alpha$. Note that the phase angles are related to Chern number up to modulus 2. 2e4 inference steps are used.}
\label{tab:int_chern}
\end{table}

\begin{figure}[H]
\centering
\includegraphics[width=1\textwidth]{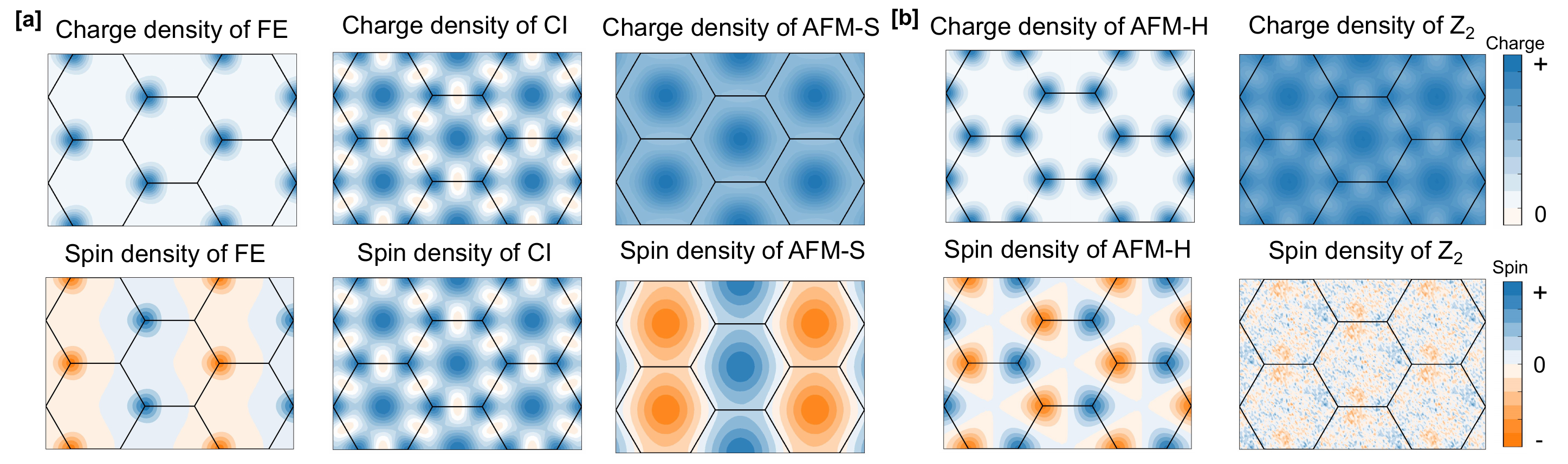}
\caption{\label{fig:density_int} Densities of integer topological insulator. \textbf{a}, charge and spin density of calculated phases at $n=-1$, including ferroelectric state, Chern insulator and antiferromagnetic stripe state.  Spin density of the $S_z=0$ ferroelectric state is plotted and shows an antiferromagnetic strip pattern, whose energy is nearly degenerated with the spin-polarized ($S_z=1$) ferroelectric state. \textbf{b}, charge and spin density of calculated phases at $n=-2$, including antiferromagnetic honeycomb state and $Z_2$ topological insulator. Note that spin density of $Z_2$ topological insulator is nearly vanishing.}
\end{figure}

Charge density $n_\uparrow+n_\downarrow$ and spin density $n_\uparrow-n_\downarrow$ of calculated phases are given in Fig.~\ref{fig:density_int}.

\section{Fractional Filling Result}
In our ED calculations, Madelung constant $v_M$, which accounts for interactions between image charges, is omitted. This term is included in QMC simulations. When comparing the energies from ED and QMC, we reformulate the ED energies based on the raw values as   $E_{\rm ED}=E^{\rm raw}_{\rm ED}+\frac{N}{2}v_M$, where $N$ denotes electron numbers.

\begin{figure}[H]
\centering
\includegraphics[width=0.7\textwidth]{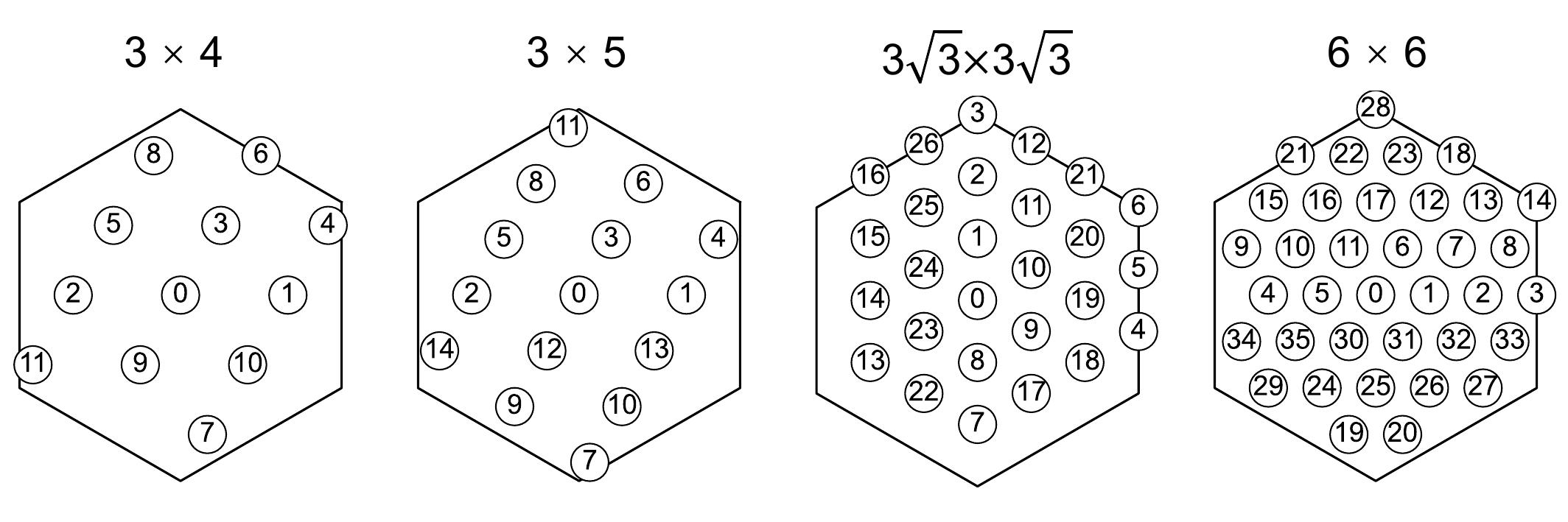}
\caption{\label{fig:index} Momentum index. Inset numbers denote momentum indexes of corresponding quantum states.
}
\end{figure}

Momentum index of our simulation supercell is plotted in Fig.~\ref{fig:index}.
For $n=-2/3$, $3\times 4$ supercell is selected. The energies at each momentum are given in Tab.~\ref{tab:frac_en}.
\begin{table}[H]
  \centering
\begin{tabularx}{0.95\textwidth}{Y|Y|Y|Y|Y}
  \hline
  \hline
      \multicolumn{5}{c}{  Total energy (meV) } \\
      \hline
    $k_x+3k_y$ & One-band ED & Two-band ED & Three-band ED & DeepSolid \\
    \hline
    0 & -377.05569 & -383.70313& -387.14490& -388.9755(5)\\
    1 & -376.81974 & -384.01252& -387.06221& -389.0973(4)\\
    2 & -376.81973 & -384.01251& -387.06221& -389.1040(4)\\
    3 & -374.62671 & -382.42348& -385.87810& -388.1034(5)\\
    4 & -374.72934 & -382.73929& -385.98177& -388.02310(4)\\
    5 & -374.38219 & -382.52770& -386.01334& -387.9889(5)\\
    6 & -375.05897 & -382.68649& -386.29079& -388.2679(5)\\
    7 & -375.19204 & -382.63668& -386.34853& -388.1190(5)\\
    8 & -375.19205 & -382.63668& -386.34853& -388.1484(5)\\
    9 & -374.62671 & -382.42348& -385.87810& -388.0786(5)\\
    10 & -374.25986 & -382.52770 & -386.01335 & -387.9749(5)\\
    11 & -374.72934 & -382.73930 & -385.98178 & -387.9981(5)\\
  
  \hline
  \hline
\end{tabularx}
\caption{Calculated total energy at fractional fillings $n=-2/3$. $3\times4$ supercell is used. Twist angle $\theta=2^\circ$. ED results are listed for comparison. Madelung constant $v_M=-17.235714\ {\rm meV}$ is included. }
\label{tab:frac_en}
\end{table}

\begin{figure}[H]
\centering
\includegraphics[width=0.95\textwidth]{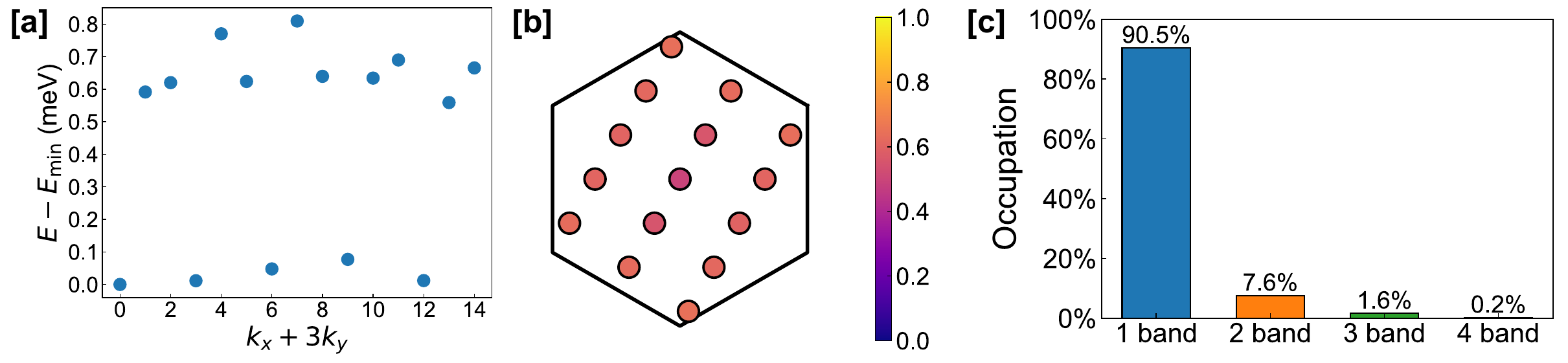}
\caption{\label{fig:3_5} Results of filling $n=-3/5$. \textbf{a}, calculated energy spectrum. $3\times5$ supercell is employed and $\theta=2^\circ$. Ground states are fivefold degenerated and center-of-mass momenta are consistent with generalized Pauli principle.
\textbf{b}, one-body density matrix averaged over five degenerated states.
\textbf{c}, band-mixing analysis.
}
\end{figure}

For $n=-3/5$, $3\times 5$ supercell is selected. And the energies of each momentum are given in Tab.~\ref{tab:frac_en_35} and Fig.~\ref{fig:3_5}.
\begin{table}[H]
  \centering
\begin{tabularx}{0.95\textwidth}{Y|Y|Y}
  \hline
  \hline
      \multicolumn{3}{c}{  Total energy (meV) } \\
      \hline
    $k_x+3k_y$ & Two-band ED & DeepSolid \\
    \hline
    0 & -424.44365 & -428.928(1) \\
    1 & -423.70821 & -428.338(1) \\
    2 & -423.70821 & -428.309(1) \\
    3 & -424.38104 & -428.9175(9) \\
    4 & -423.54971 & -428.159(1) \\
    5 & -423.59749 & -428.305(1) \\
    6 & -424.30492 & -428.881(1) \\
    7 & -423.58448 & -428.1194(8) \\
    8 & -423.57328 & -428.2894(9) \\
    9 & -424.30492 & -428.852(1) \\
    10 & -423.57328 & -428.295(1) \\
    11 & -423.58448 & -428.239(1) \\
    12 & -424.38104 & -428.917(1) \\
    13 & -423.59749 & -428.370(2) \\
    14 & -423.54971 & -428.264(4) \\
  
  \hline
  \hline
\end{tabularx}
\caption{Calculated total energy at fractional fillings $n=-3/5$. $3\times5$ supercell is used. Twist angle $\theta=2^\circ$. ED results are listed for comparison. Madelung constant $v_M=-15.10449\ {\rm meV}$ is included. }
\label{tab:frac_en_35}
\end{table}

For $n=-1/3$, $3\sqrt{3}\times3\sqrt{3}$ and $6\times 6$ supercells are used. 
The energies of $3\sqrt{3}\times 3\sqrt{3}$ supercell are given in Tab.~\ref{tab:frac_en_sqrt3}.

\begin{table}[H]
  \centering
\begin{tabularx}{0.95\textwidth}{Y|Y|Y|Y|Y|Y}
  \hline
  \hline
      \multicolumn{6}{c}{  Total energy (meV) } \\
      \hline
    Momentum  & One-band ED & DeepSolid & Momentum  &  One-band ED & DeepSolid \\
    \hline
    0 & -387.19570 & -389.4147(6) & 14 & -384.36684 & -387.694(1) \\
    1 & -384.34012 & -388.1231(8) & 15 & -384.47890 & -388.1620(8) \\
    2 & -384.47890 & -388.1261(9) & 16 & -384.99761 & -388.3358(9) \\
    3 & -385.75033 & -389.2814(7) & 17 & -384.28700 & -387.8082(9) \\
    4 & -384.99761 & -388.3578(9) & 18 & -384.47890 & -388.0801(9) \\
    5 & -384.99761 & -388.2633(8) & 19 & -384.36684 & -387.8415(9) \\
    6 & -385.75033 & -389.2935(7) & 20 & -384.47890 & -388.1152(8) \\
    7 & -384.47890 & -388.1701(7) & 21 & -384.99761 & -388.3212(8) \\
    8 & -384.34012 & -387.914(1) & 22 & -384.36684 & -387.790(1) \\
    9 & -384.34012 & -388.1211(9) & 23 & -384.34012 & -388.1611(9) \\
    10 & -384.34012 & -388.1540(8) & 24 & -384.34012 & -388.1606(8) \\
    11 & -384.36684 & -387.7936(9) & 25 & -384.36684 & -387.8300(9) \\
    12 & -384.99761 & -388.3149(9) & 26 & -384.99761 & -388.184(1)\\
    13 & -384.47890 & -388.1635(8) & & & \\
  
  \hline
  \hline
\end{tabularx}
\caption{Calculated total energy at fractional fillings $n=-1/3$. $3\sqrt{3}\times3\sqrt{3}$ supercell is used. Twist angle $\theta=2^\circ$. ED results are listed for comparison. Madelung constant $v_M=-11.57837\ {\rm meV}$ is included. }
\label{tab:frac_en_sqrt3}
\end{table}

The energies of $6\times 6$ supercell are given in Tab.~\ref{tab:frac_en_66}.
\begin{table}[H]
  \centering
\begin{tabularx}{0.95\textwidth}{Y|Y|Y|Y}
  \hline
  \hline
      \multicolumn{4}{c}{  Total energy (meV) } \\
      \hline
    $k_x+6k_y$  & DeepSolid & $k_x+6k_y$  & DeepSolid \\
    \hline
    0 & -518.8027(5) & 18 & -517.5582(8)\\
    1 & -517.4905(7) & 19 & -517.5397(8)\\
    2 & -517.1435(8) & 20 & -517.1141(8)\\
    3 & -517.5355(7) & 21 & -517.5334(7)\\
    4 & -517.2232(8) & 22 & -517.1038(9)\\
    5 & -517.4797(8) & 23 & -517.5115(8)\\
    6 & -517.4866(8) & 24 & -517.1930(9)\\
    7 & -517.3063(7) & 25 & -517.2896(7)\\
    8 & -517.5453(7) & 26 & -517.1011(8)\\
    9 & -517.6046(8) & 27 & -517.4946(8)\\
    10 & -517.2950(7) & 28 & -518.7395(5)\\
    11 & -517.4929(7) & 29 & -517.1476(9)\\
    12 & -517.1040(7) & 30 & -517.4764(7)\\
    13 & -517.5490(9) & 31 & -517.3999(8)\\
    14 & -518.7554(5) & 32 & -517.3086(7)\\
    15 & -517.5403(8) & 33 & -517.5435(8)\\
    16 & -517.1413(9) & 34 & -517.5344(7)\\
    17 & -517.3094(7) & 35 & -517.3166(8)\\
  
  \hline
  \hline
\end{tabularx}
\caption{Calculated total energy at fractional fillings $n=-1/3$. $6\times6$ supercell is used. Twist angle $\theta=2^\circ$. Madelung constant $v_M=-10.027165\ {\rm meV}$ is included. }
\label{tab:frac_en_66}
\end{table}

\bibliography{supplement}